%% file: cds.tex
\newif\ifnature
\naturefalse

\documentclass[ALICE,manyauthors]{cernphprep}
\pdfoutput=1
\usepackage{graphicx}
\usepackage{hyperref}
\usepackage{multirow}
\usepackage{lineno}
\usepackage[normalem]{ulem}
\usepackage[comma,square,numbers,sort&compress]{natbib}
\usepackage{tabularx,booktabs}
\newcolumntype{Y}{>{\centering\arraybackslash}X}
\AtBeginDocument{
\heavyrulewidth=.08em
\lightrulewidth=.05em
\cmidrulewidth=.03em
\belowrulesep=.65ex
\belowbottomsep=0pt
\aboverulesep=.4ex
\abovetopsep=0pt
\cmidrulesep=\doublerulesep
\cmidrulekern=.5em
\defaultaddspace=.5em
}
\input{commands}

\begin{document}

\PHyear{2016}
\PHnumber{153}
\PHdate{11 June}

\begin{titlepage}

\title{Enhanced production of multi-strange hadrons\\in high-multiplicity proton-proton collisions}
\ShortTitle{Enhanced production of multi-strange hadrons in high-multiplicity pp}
\author{ALICE Collaboration}

\begin{abstract}
  At sufficiently high temperature and energy density, nuclear matter undergoes a transition to a phase in which quarks and gluons are not confined: the Quark-Gluon Plasma (QGP)~\cite{Shuryak:1980tp}.
  Such an extreme state of strongly-interacting QCD (Quantum Chromo-Dynamics) matter is produced in the laboratory with high-energy collisions of heavy nuclei, where an enhanced production of strange hadrons is observed~\cite{Andersen:1999ym,Afanasiev:2002he,Antinori:2004ee,Abelev:2007xp,ABELEV:2013zaa}.
  Strangeness enhancement, originally proposed as a signature of QGP formation in nuclear collisions~\cite{Koch:1986ud}, is  more pronounced for multi-strange baryons.
  Several effects typical of heavy-ion phenomenology have been observed in high-multiplicity proton-proton (pp) collisions~\cite{Khachatryan:2010gv,Khachatryan:2016txc}.
  Yet, enhanced production of multi-strange particles has not been reported so far.
  Here we present the first observation of strangeness enhancement in high-multiplicity pp collisions.
  We find that the integrated yields of strange and multi-strange particles relative to pions increases significantly with the event charged-particle multiplicity.
  The measurements are in remarkable agreement with \pPb collision results~\cite{Abelev:2013haa,Adam:2015vsf} indicating that the phenomenon is related to the final system created in the collision.
  In high-multiplicity events strangeness production reaches values similar to those observed in \PbPb collisions, where a QGP is formed. 
\end{abstract}

\end{titlepage}

\setcounter{page}{2}

\input{paper}

\newenvironment{acknowledgement}{\relax}{\relax}
\begin{acknowledgement}
\section*{Acknowledgements}
\input{acknowledgements.tex}    
\end{acknowledgement}

\bibliographystyle{utphys}   
\bibliography{biblio}

\newpage
\appendix
\input{methods}

\section{The ALICE Collaboration}
\label{app:collab}
\input{Alice_Authorlist_2016-May-30.tex}  
\end{document}

\end{document}

%% file: commands.tex
\usepackage{xspace}
\usepackage{color}
\usepackage{xcolor}
\usepackage{rotating}
\definecolor{light-gray}{gray}{0.8}

\newcommand{\pt}{\ensuremath{p_{\rm T}}\xspace}
\newcommand{\mt}{\ensuremath{m_{\rm T}}\xspace}
\newcommand{\sqrts}{\ensuremath{\sqrt{s}}\xspace}
\newcommand{\sqrtsNN}{\ensuremath{\sqrt{s_{\rm NN}}}\xspace}
\newcommand{\nch}{\ensuremath{N_{\rm ch}}\xspace}
\newcommand{\dnchdeta}{\ensuremath{{\rm d}N_{\rm ch}/{\rm d}\eta}\xspace}

\newcommand{\etaless}[1]{\ensuremath{\left|\eta\right| < #1}\xspace}
\newcommand{\yless}[1]{\ensuremath{\left|y\right| < #1}\xspace}

\newcommand{\GeVc}{GeV/$c$\xspace}
\newcommand{\pp}{\ensuremath{\rm pp}\xspace}
\newcommand{\pPb}{p--Pb\xspace}
\newcommand{\PbPb}{Pb--Pb\xspace}

\newcommand{\inelgtzero}{INEL$>$0\xspace}

\newcommand{\average}[1]{\ensuremath{\langle #1 \rangle}\xspace}

\newcommand{\pPiplus}{\ensuremath{{\pi}^{+}}\xspace}
\newcommand{\pPiplusNS}{\ensuremath{{\pi}^{+}}}
\newcommand{\pPiminus}{\ensuremath{{\pi}^{-}}\xspace}
\newcommand{\pPiminusNS}{\ensuremath{{\pi}^{-}}}
\newcommand{\sPi}{\ensuremath{{\pi}}\xspace}

\newcommand{\pKplusNS}{\ensuremath{{\rm K}^{+}}}

\newcommand{\pKminusNS}{\ensuremath{{\rm K}^{-}}}
\newcommand{\sProton}{\ensuremath{\rm p}\xspace}
\newcommand{\pProton}{\ensuremath{\rm p}\xspace}
\newcommand{\pProtonNS}{\ensuremath{\rm p}}
\newcommand{\apProton}{\ensuremath{\overline{\rm p}}\xspace}
\newcommand{\apProtonNS}{\ensuremath{\overline{\rm p}}}
\newcommand{\sPr}{\ensuremath{\rm p}\xspace}
\newcommand{\sKzero}{\ensuremath{2{\rm K}^{0}_{S}}\xspace}
\newcommand{\pKzero}{\ensuremath{{\rm K}^{0}_{S}}\xspace}
\newcommand{\sLambda}{\ensuremath{\Lambda}\xspace}
\newcommand{\pLambda}{\ensuremath{\Lambda}\xspace}
\newcommand{\pLambdaNS}{\ensuremath{\Lambda}}
\newcommand{\apLambda}{\ensuremath{\overline{\Lambda}}\xspace}
\newcommand{\apLambdaNS}{\ensuremath{\overline{\Lambda}}}
\newcommand{\sXi}{\ensuremath{\Xi}\xspace}
\newcommand{\pXi}{\ensuremath{\Xi^{-}}\xspace}
\newcommand{\pXiNS}{\ensuremath{\Xi^{-}}}
\newcommand{\apXi}{\ensuremath{\overline{\Xi}^{+}}\xspace}
\newcommand{\apXiNS}{\ensuremath{\overline{\Xi}^{+}}}
\newcommand{\sOmega}{\ensuremath{\Omega}\xspace}
\newcommand{\pOmega}{\ensuremath{\Omega^{-}}\xspace}
\newcommand{\pOmegaNS}{\ensuremath{\Omega^{-}}}
\newcommand{\apOmega}{\ensuremath{\overline{\Omega}^{+}}\xspace}
\newcommand{\apOmegaNS}{\ensuremath{\overline{\Omega}^{+}}}

\gdef\Adef#1{\label{LA#1}}
\gdef\Aref#1{\textsuperscript{,\ref{LA#1}}}
\gdef\Idef#1{\label{LI#1}\gdef\INTlatex{#1}}
\gdef\Irefn#1{\textsuperscript{\ref{LI#1}}}

\gdef\And{, }
\gdef\NNospacing{\labelsep=0pt\itemsep=0pt\topsep=0pt\partopsep=0pt\parskip=0pt\parsep=0pt}

\ifnature
\newenvironment{Authlist}{\vspace*{-1.5ex}\enumerate\NNospacing}
                         {\endlist\vspace*{-1.5ex}}
\fi

%% file: paper.tex

\ifnature

\textbf{
  At sufficiently high temperature and energy density, nuclear matter undergoes a transition to a phase in which quarks and gluons are not confined: the Quark-Gluon Plasma (QGP)~\cite{Shuryak:1980tp}.
  Such an extreme state of strongly-interacting QCD (Quantum Chromo-Dynamics) matter is produced in the laboratory with high-energy collisions of heavy nuclei, where an enhanced production of strange hadrons is observed~\cite{Andersen:1999ym,Afanasiev:2002he,Antinori:2004ee,Abelev:2007xp,ABELEV:2013zaa}.
  Strangeness enhancement, originally proposed as a signature of QGP formation in nuclear collisions~\cite{Koch:1986ud}, is  more pronounced for multi-strange baryons.
  Several effects typical of heavy-ion phenomenology have been observed in high-multiplicity proton-proton (pp) collisions~\cite{Khachatryan:2010gv,Khachatryan:2016txc}.
  Yet, enhanced production of multi-strange particles has not been reported so far.
  Here we present the first observation of strangeness enhancement in high-multiplicity pp collisions.
  We find that the integrated yields of strange and multi-strange particles relative to pions increases significantly with the event charged-particle multiplicity.
  The measurements are in remarkable agreement with \pPb collision results~\cite{Abelev:2013haa,Adam:2015vsf} indicating that the phenomenon is related to the final system created in the collision.
  In high-multiplicity events strangeness production reaches values similar to those observed in \PbPb collisions, where a QGP is formed. 
}

\fi


The production of strange hadrons in high-energy hadronic interactions provides a way to investigate the properties of QCD, the theory of strongly-interacting matter.
Unlike up ($u$) and down ($d$) quarks, which form ordinary matter, strange ($s$) quarks are not present as valence quarks in the initial state, yet they are sufficiently light to be abundantly created during the course of the collisions.
In the early stages of high energy collisions, strangeness is produced in hard (perturbative) 2$\,\to\,$2 partonic scattering processes by flavour creation ($gg~\to~s\bar{s}$, $q\bar{q}~\to~s\bar{s}$) and flavour excitation ($gs~\to~gs$, $qs~\to~qs$).
Strangeness is also created during the subsequent partonic evolution via gluon splittings ($g~\to~s\bar{s}$). These processes tend to dominate the production of high transverse momentum (\pt) strange hadrons. At low \pt non perturbative processes dominate the production of strange hadrons. In string fragmentation models the production of strange hadrons is generally suppressed relative to hadrons containing only light quarks, as the strange quark is heavier than up and down quarks.
The amount of strangeness suppression in elementary ($e^+e^-$ and pp) collisions is an important parameter in Monte Carlo (MC) models.
For this reason, measurements of strange hadron production place constraints on these models.

The abundances of strange particles relative to pions in heavy-ion collisions from top RHIC (Relativistic Heavy-Ion Collider) to LHC (Large Hadron Collider) energies do not show a significant dependence on either the initial volume (collision centrality) or the initial energy density (collision energy). With the exception of the most peripheral collisions, particle ratios are found to be compatible with those of a hadron gas in thermal and chemical equilibrium and can be described using a grand canonical statistical model~\cite{Cleymans:2006xj,Andronic:2008gu}.
In peripheral collisions, where the overlap of the colliding nuclei becomes very small, the relative yields of strange particles to pions decrease and tend toward those observed in pp collisions, for which a statistical-mechanics approach can also be applied~\cite{Hagedorn:1967ua,Becattini:1997rv}.
Extensions of a pure grand-canonical description of particle production, like statistical models implementing strangeness canonical suppression~\cite{Redlich:2001kb} and core-corona superposition~\cite{Becattini:2008yn,Aichelin:2008mi} models, can effectively produce a suppression of strangeness production in small systems. However, the microscopic origin of enhanced strangeness production is not known, and the measurements presented in this Letter may contribute to its understanding.
Several effects, like azimuthal correlations
and mass-dependent hardening of \pt distributions, which in nuclear collisions are typically attributed to the formation of a strongly-interacting quark-gluon medium, have been observed in high-multiplicity pp and proton-nucleus collisions at the LHC~\cite{Khachatryan:2010gv,CMS:2012qk,Abelev:2012ola,Aad:2012gla,Aad:2013fja,Chatrchyan:2013nka,Abelev:2013haa,ABELEV:2013wsa,Adam:2015vsf,Khachatryan:2016yru,Khachatryan:2016txc}. Yet, enhanced production of strange particles as a function of the charged-particle multiplicity density (\dnchdeta)
has so far not been observed in pp collisions.
The study of pp collisions at high multiplicity is thus of considerable interest as it opens the exciting possibility of a microscopic understanding of phenomena known from nuclear reactions.

In this Letter, we present the multiplicity dependence of the production of primary strange (\pKzero, \pLambda, \apLambda) and multi-strange (\pXi, \apXi, \pOmega, \apOmega) hadrons in pp collisions at the centre-of-mass energy of \sqrts~=~7 TeV.
Primary particles are defined as all particles created in the collisions, except those coming from weak decays of light-flavour hadrons and of muons.
The measurements have been performed at midrapidity\footnote{The particle rapidity is defined as $y = \frac{1}{2} \ln \left (\frac{E + p_{z}c}{E - p_{z}c} \right)$, where $E$ is the energy and $p_{z}$ is the component of momentum along the beam axis.}, \yless{0.5}, with the ALICE detector~\cite{Aamodt:2008zz} at the LHC.
Similar measurements of the multiplicity and centrality dependence of strange and multi-strange hadron production have been performed by ALICE in proton-lead (\pPb) collisions at a centre-of-mass energy per nucleon pair \sqrtsNN~=~5.02 TeV~\cite{Abelev:2013haa,Adam:2015vsf} and in lead-lead (\PbPb) collisions at \sqrtsNN~=~2.76 TeV~\cite{Abelev:2013xaa,ABELEV:2013zaa}.
The measurements reported here have been obtained in pp collisions at \sqrts~=~7 TeV for events having at least one charged particle produced in the pseudorapidity\footnote{The particle pseudorapidity is defined as $\eta = -\ln \left( tan \frac{\theta}{2} \right)$, where $\theta$ is the angle with respect to the beam axis.} interval \etaless{1} (\inelgtzero), corresponding to about 75\% of the total inelastic cross-section. In order to study the multiplicity dependence of strange and multi-strange hadron production, the sample is divided into event classes based on the total ionisation energy deposited in the forward detectors, covering the pseudorapidity regions $2.8~<~\eta~<~5.1$ and $-3.7~<~\eta~<~-1.7$.


\ifnature
\begin{figure}[pt]
\else
\begin{figure}[t]
\fi
    \centering
    \ifnature
    \includegraphics[width=\linewidth]{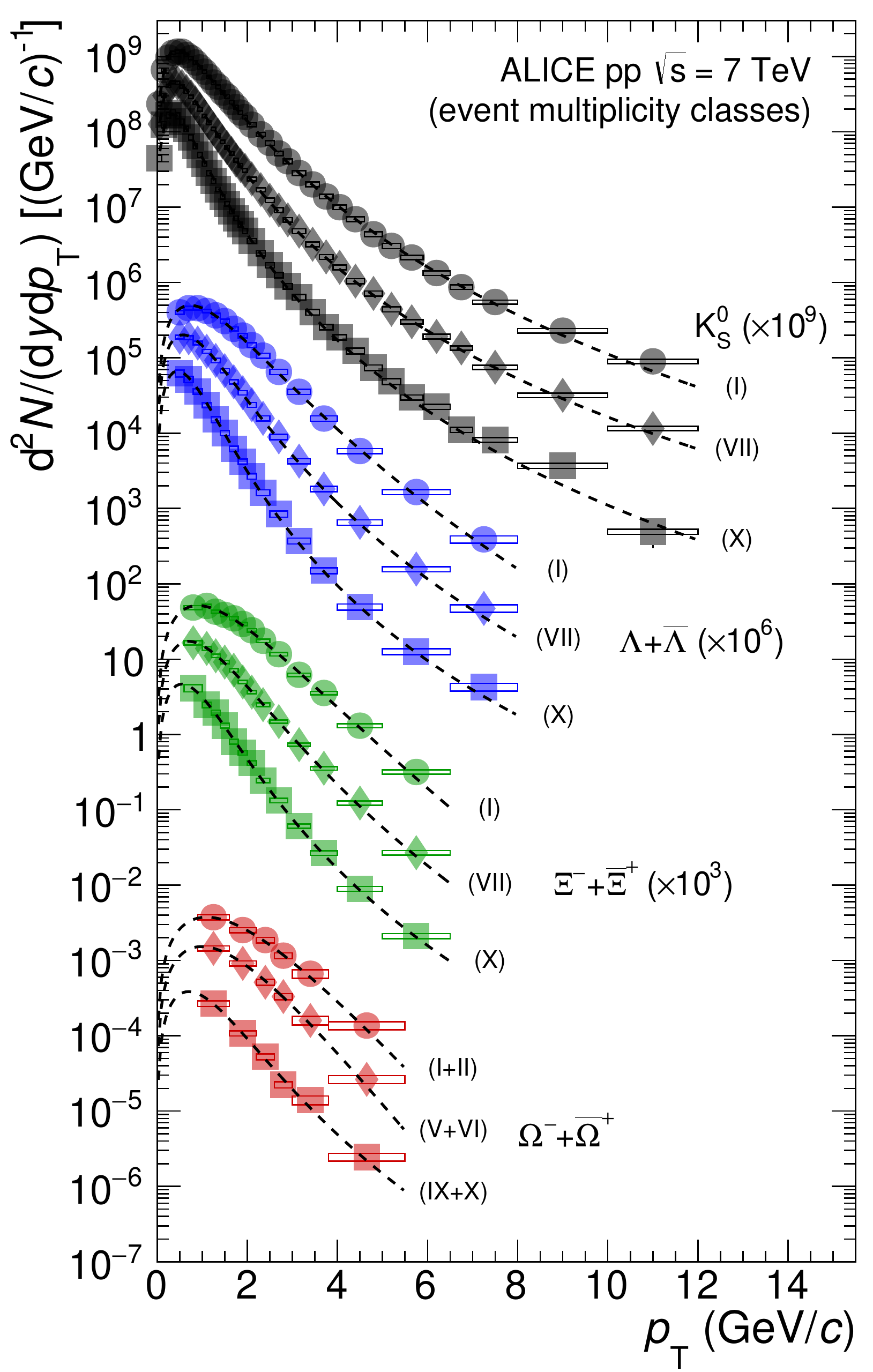}
    \else
    \includegraphics[width=0.5\linewidth]{spectra.pdf}
    \fi
\caption{{\ifnature\bf\fi \pt-differential yields of $\pKzero$, $\pLambda+\apLambda$, $\pXi+\apXi$ and $\pOmega+\apOmega$ measured in \yless{0.5}.} The results are shown for a selection of event classes, indicated by roman numbers in brackets, with decreasing multiplicity. The error bars show the statistical uncertainty, whereas the empty boxes show the total systematic uncertainty. The data are scaled by different factors to improve the visibility. The dashed curves represent Tsallis-L\'evy fits to each individual distribution to extract integrated yields. The indicated uncertainties all represent standard deviations.}
  \label{fig:spectra}
\end{figure}

Particle/antiparticle production yields are identical within uncertainties. The \pt distributions of \pKzero, $\pLambda+\apLambda$, $\pXi+\apXi$ and $\pOmega+\apOmega$ (in the following denoted as \pKzero, \sLambda, \sXi and \sOmega) are shown in Figure~\ref{fig:spectra} for a selection of event classes with progressively decreasing \average{\dnchdeta}. The mean pseudorapidity densities of primary charged particles \average{\dnchdeta} are measured at midrapidity, \etaless{0.5}.
The \pt spectra become harder as the multiplicity increases, with the hardening being more pronounced for higher mass particles.
A similar observation was reported for \pPb collisions~\cite{Abelev:2013haa} where this and several other features common with \PbPb collisions are consistent with the appearance of collective behavior at high-multiplicity~\cite{Khachatryan:2010gv,CMS:2012qk,Abelev:2012ola,Aad:2012gla,Aad:2013fja,Chatrchyan:2013nka,Adam:2015vsf}.
In heavy-ion collisions these observations are successfully described by models based on relativistic hydrodynamics. In this framework, the \pt distributions are determined by particle emission from a collectively expanding thermal source~\cite{Heinz:2004qz}. 
The blast-wave model~\cite{Schnedermann:1993ws} is employed to analyse the spectral shapes of \pKzero, \sLambda and \sXi in the common highest multiplicity class (class I). A simultaneous fit to all particles is performed following the approach discussed in~\cite{Abelev:2013haa} in the \pt ranges 0--1.5, 0.6--2.9 and 0.6--2.9~\GeVc, for \pKzero, \sLambda and \sXi, respectively. The best-fit describes the data to better than 5\% in the respective fit ranges, consistent with particle production from a thermal source at temperature $T_{\rm fo}$ expanding with a common transverse velocity \average{\beta_{\rm T}}. The resulting parameters, $T_{\rm fo}$~=~163~$\pm$~10~MeV and \average{\beta_{\rm T}}~=~0.49~$\pm$~0.02, are remarkably similar to the ones obtained in \pPb collisions for an event class with comparable \average{\dnchdeta}~\cite{Abelev:2013haa}.

\ifnature
\begin{figure}[pt]
\else
\begin{figure}[t]
\fi
        \centering
        \ifnature
        \includegraphics[width=\linewidth]{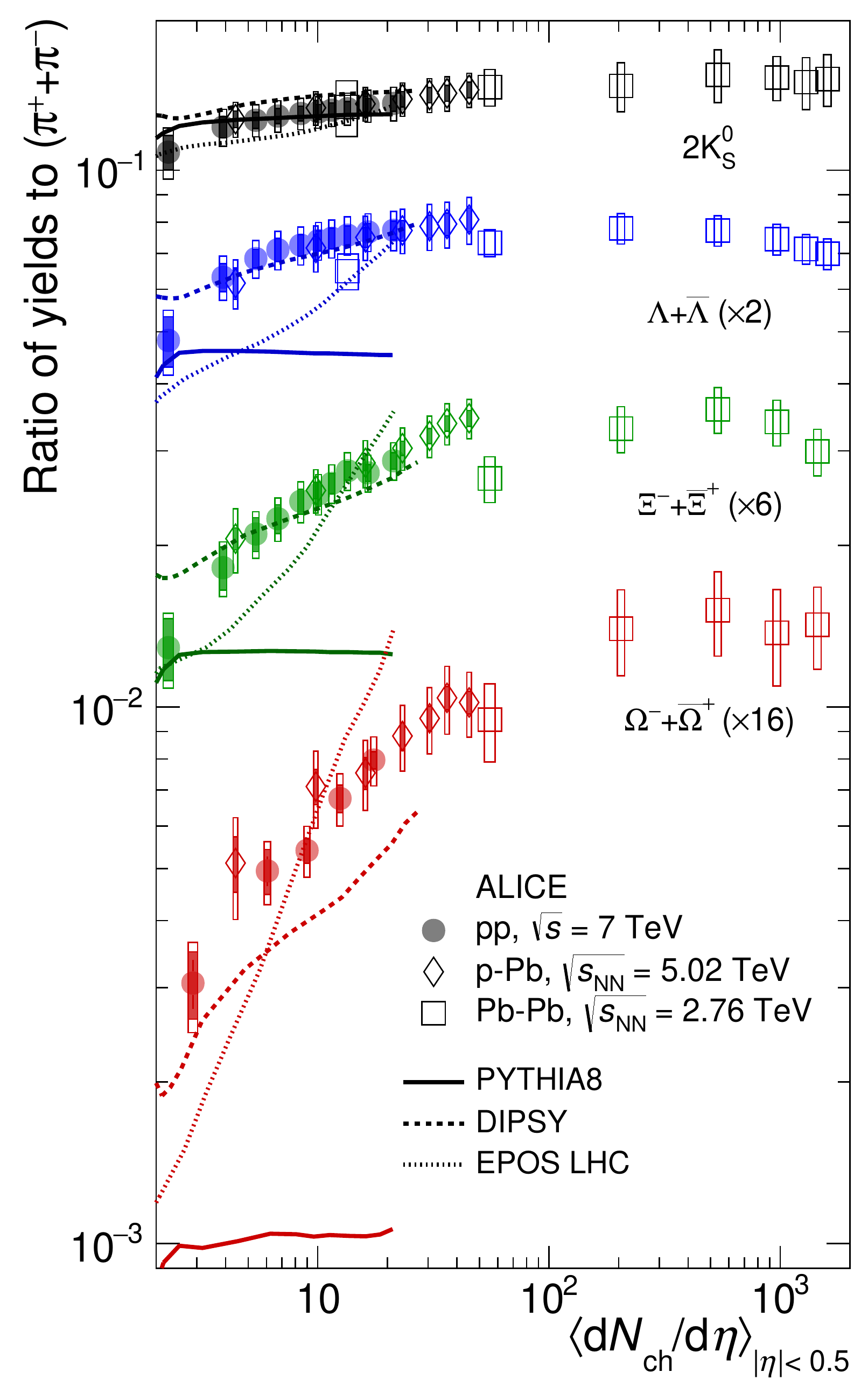}
        \else
        \includegraphics[width=0.5\linewidth]{ratios.pdf}
        \fi
\caption{{\ifnature\bf\fi \pt-integrated yield ratios to pions ($\pPiplus+\pPiminus$) as a function of \average{\dnchdeta} measured in \yless{0.5}.}
    The error bars show the statistical uncertainty, whereas the empty and dark-shaded boxes show the total systematic uncertainty and the contribution uncorrelated across multiplicity bins, respectively.
    The values are compared to calculations from MC models~\cite{Sjostrand:2007gs,Pierog:2013ria,Bierlich:2015rha} and to results obtained in \pPb and \PbPb collisions at the LHC~\cite{Abelev:2013haa,ABELEV:2013zaa,Adam:2015vsf}. For \PbPb results the ratio $2 \Lambda$~/~($\pPiplus+\pPiminus$) is shown. The indicated uncertainties all represent standard deviations.}
  \label{fig:ratios}
\end{figure}

The \pt-integrated yields are computed from the data in the measured ranges and using extrapolations to the unmeasured regions. In order to extrapolate to the unmeasured region, the data were fitted with a Tsallis-L\'evy~\cite{Abelev:2013haa} parametrization, which gives the best description of the individual spectra for all particles and all event classes over the full \pt range (Figure~\ref{fig:spectra}). Several other fit functions 
(Boltzmann, \mt-exponential, \pt-exponential, blast-wave, Fermi-Dirac, Bose-Einstein) are employed to estimate the corresponding systematic uncertainties. The fraction of the extrapolated yield for the highest(lowest) multiplicity event class is about 10(25)\%, 16(36)\%, 27(47)\% for \sLambda, \sXi and \sOmega, respectively, and is negligible for \pKzero. The uncertainty on the extrapolation amounts to about 2(6)\%, 3(10)\%, 4(13)\% of the total yield for \sLambda, \sXi and \sOmega, respectively, and it is negligible for \pKzero. The total systematic uncertainty on the \pt-integrated yields amounts to 5(9)\%, 7(12)\%, 6(14)\% and 9(18)\% for \pKzero, \sLambda, \sXi and \sOmega, respectively. A significant fraction of this uncertainty is common to all multiplicity classes and it is estimated to be about 5\%, 6\%, 6\% and 9\% for \pKzero, \sLambda, \sXi and \sOmega, respectively.
In Figure~\ref{fig:ratios}, the ratios of the yields of \pKzero, \sLambda, \sXi and \sOmega to the pion ($\pi^{+}+\pi^{-}$) yield as a function of \average{\dnchdeta} are compared to \pPb and \PbPb results at the LHC~\cite{Abelev:2013haa,Adam:2015vsf,ABELEV:2013zaa}.
A significant enhancement of strange to non-strange hadron production is observed with increasing particle multiplicity in pp collisions. The behaviour observed in pp collisions resembles that of \pPb collisions at a slightly lower centre-of-mass energy~\cite{Adam:2015vsf}, in terms of both the values of the ratios and their evolution with multiplicity. As no significant dependence on the centre-of-mass energy is observed at the LHC for inclusive inelastic collisions, the origin of strangeness production in hadronic collisions is apparently driven by the characteristics of the final state rather than by the collision system or energy. At high multiplicity, the yield ratios 
reach values similar to the ones observed in \PbPb collisions, where no significant change with multiplicity is observed beyond an initial slight rise.
Note that the final-state average charged-particle density \average{\dnchdeta}, which changes by over three orders of magnitude from low-multiplicity pp to central \PbPb, will in general be related to different underlying physics in the various reaction systems. For example, under the assumption that the initial reaction volume in both \pp and \pPb is determined mostly by the size of the proton, \average{\dnchdeta} could be used as a proxy for the initial energy density. In \PbPb collisions, on the other hand, both the overlap area as well as the energy density could increase with \average{\dnchdeta}. Nonetheless, it is a non-trivial observation that particle ratios in pp and \pPb are identical at the same \dnchdeta, representing an indication that the final-state particle density might indeed be a good scaling variable between these two systems.

\begin{figure}[t]
    \centering
    \ifnature
    \includegraphics[width=\linewidth]{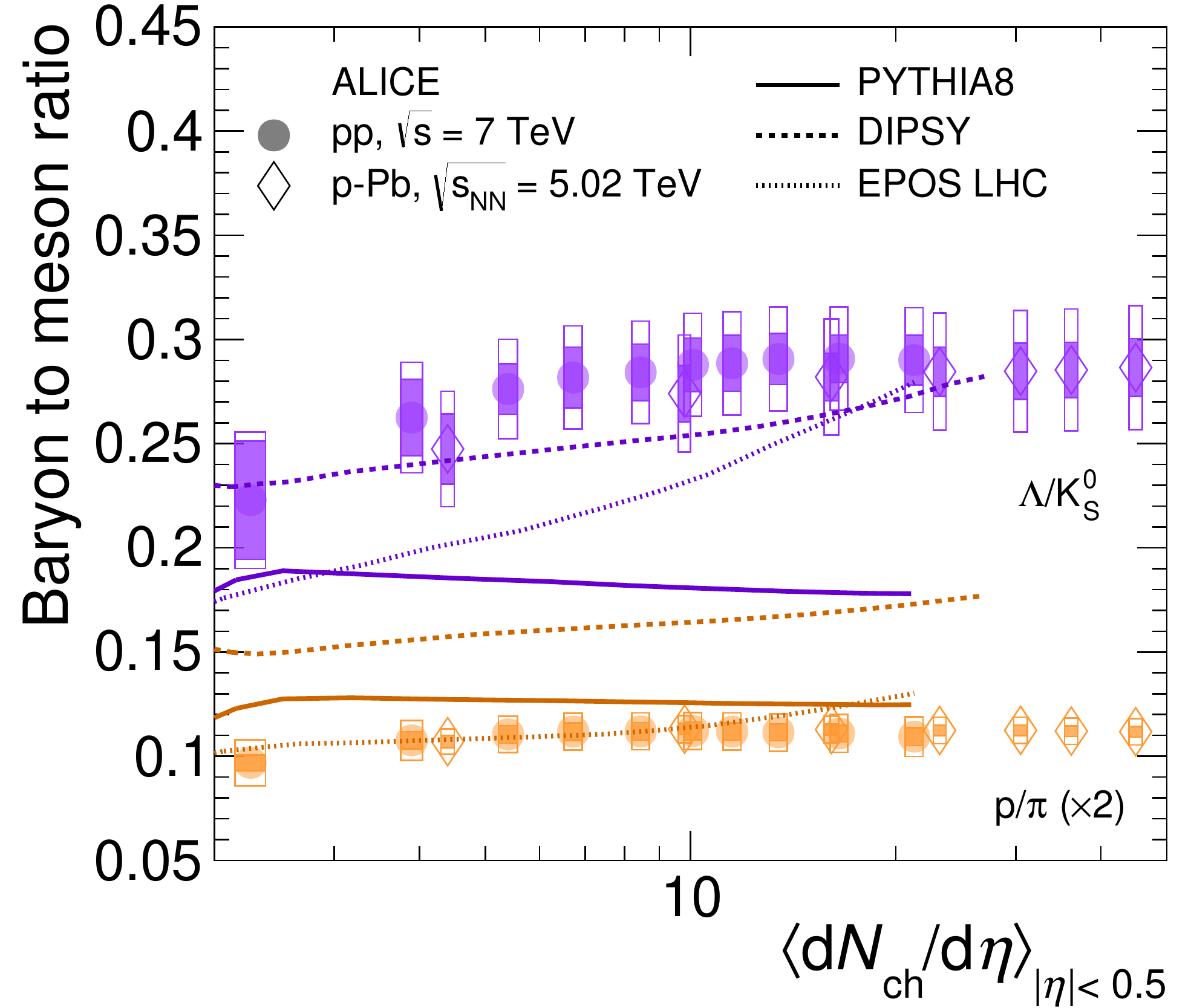}
    \else
    \includegraphics[width=0.5\linewidth]{bmratios.pdf}
    \fi    
\caption{{\ifnature\bf\fi Particle yield ratios \sLambda/\pKzero = ($\pLambda+\apLambda$)/\sKzero and \sPr/\sPi = ($\pProton+\apProton$)/($\pPiplus+\pPiminus$) as a function of \average{\dnchdeta}.}
      The yield ratios are measured in the rapidity interval \yless{0.5}.
      The error bars show the statistical uncertainty, whereas the empty and dark-shaded boxes show the total systematic uncertainty and the contribution uncorrelated across multiplicity bins, respectively.
      The values are compared to calculations from MC models~\cite{Sjostrand:2007gs,Pierog:2013ria,Bierlich:2015rha}
      in pp collisions at \sqrts = 7 TeV and to results obtained in \pPb collisions at the LHC~\cite{Abelev:2013haa}. The indicated uncertainties all represent standard deviations.}
  \label{fig:bmratios}
  \end{figure}  

Figure~\ref{fig:bmratios} shows that the yield ratios \sLambda/\pKzero = ($\pLambda+\apLambda$)/\sKzero and \sProton/\sPi = ($\pProton+\apProton$)/($\pPiplus+\pPiminus$) do not change significantly with multiplicity, demonstrating that the observed enhanced production rates of strange hadrons with respect to pions is not due to the difference in the hadron masses.
The results in Figures~\ref{fig:ratios} and~\ref{fig:bmratios} are compared to calculations from MC models commonly used for pp collisions at the LHC: PYTHIA8~\cite{Sjostrand:2007gs}, EPOS LHC~\cite{Pierog:2013ria} and DIPSY~\cite{Bierlich:2015rha}.
The kinematic domain and the multiplicity selections are the same for MC and data, namely, dividing the \inelgtzero sample into event classes based on the total charged-particle multiplicity in the forward region.
The observation of a multiplicity-dependent enhancement of the production of strange hadrons along with the constant production of protons relative to pions cannot be simultaneously reproduced by any of the MC models commonly used at the LHC.
The model which describes the data best, DIPSY, is a model where interaction between gluonic strings is allowed to form ``color ropes'' which are expected to produce more strange particles and baryons.

\begin{figure}[t]
    \centering
    \ifnature
    \includegraphics[width=\linewidth]{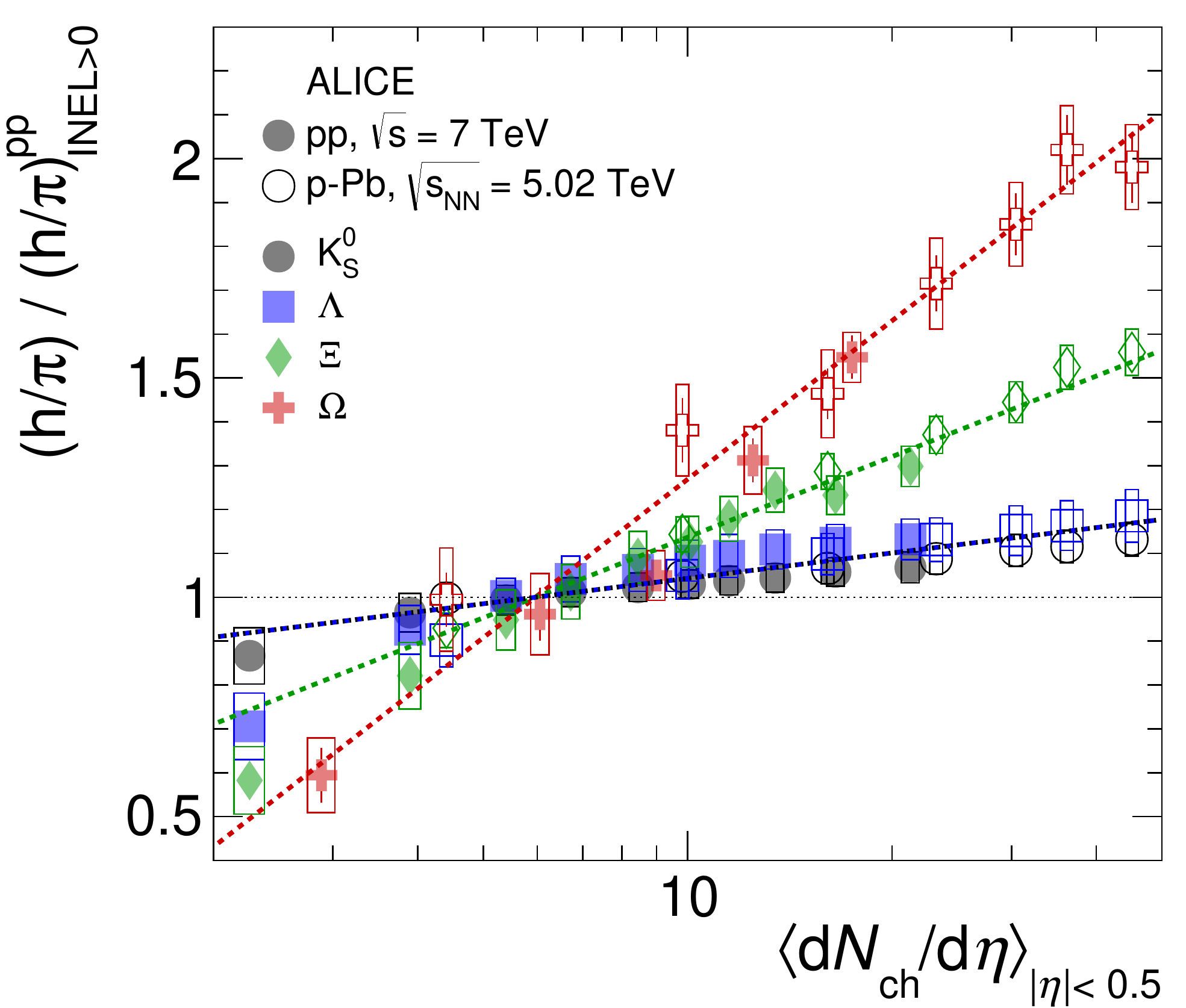}
    \else
    \includegraphics[width=0.5\linewidth]{scaling.pdf}
    \fi
\caption{{\ifnature\bf\fi Particle yield ratios to pions normalised to the values measured in the inclusive \inelgtzero pp sample.} The results are shown for pp and \pPb collisions, both normalised to the inclusive \inelgtzero pp sample. The error bars show the statistical uncertainty. The common systematic uncertainties cancel in the double-ratio. The empty boxes represent the remaining uncorrelated uncertainties. The lines represent a simultaneous fit of the results with the empirical scaling formula in Equation~\ref{eq:scaling}. The indicated uncertainties all represent standard deviations.}
  \label{fig:scaling}
\end{figure}

To illustrate the evolution of the production of strange hadrons with multiplicity, Figure~\ref{fig:scaling} presents the yield ratios to pions divided by the values measured in the inclusive \inelgtzero pp sample, both for pp and \pPb collisions. The observed multiplicity-dependent enhancement with respect to the \inelgtzero sample follows a hierarchy determined by the
hadron strangeness.
We have attempted to describe the observed strangeness hierarchy by fitting the data presented in Figure~\ref{fig:scaling} an the empirical function of the form
\begin{equation}
  \frac{(h/\pi)}{(h/\pi)^{\rm pp}_{\rm INEL>0}}  = 1~+~a~{\rm S}^{b}~\log{\left[\frac{\average{\dnchdeta}}{\average{\dnchdeta}^{\rm pp}_{\rm INEL>0}}\right]},
  \label{eq:scaling}
\end{equation}
where S is the number of strange or anti-strange valence quarks in the hadron, (h/$\pi$)$^{\rm pp}_{\rm INEL>0}$ and \ifnature\else\linebreak\fi$\average{\dnchdeta}^{\rm pp}_{\rm INEL>0}$ are the measured hadron-to-pion ratio and the charged-particle multiplicity density in \inelgtzero pp collisions, respectively, and $a$ and $b$ are free parameters. The fit describes the data well, yielding $a~=~0.083~\pm~0.006$, $b~=~1.67~\pm~0.09$, with a $\chi^{2}$/ndf of 0.66.


In summary, we have presented the multiplicity dependence of the production of primary strange (\pKzero, \pLambda, \apLambda) and multi-strange (\pXi, \apXi, \pOmega, \apOmega) hadrons in pp collisions at \sqrts = 7 TeV. The results are obtained as a function of \average{\dnchdeta} measured at midrapidity for event classes selected on the basis of the total charge deposited in the forward region.
The \pt spectra become harder as the multiplicity increases.
The mass and multiplicity dependences of the spectral shapes are reminiscent of the patterns seen in \pPb and \PbPb collisions at the LHC, which can be understood assuming a collective expansion of the system in the final state.
The data show for the first time in pp collisions that the \pt-integrated yields of strange and multi-strange particles relative to pions increase significantly with multiplicity. These particle ratios are similar to those found in \pPb collisions at the same multiplicity densities~\cite{Adam:2015vsf}. 
The observed enhancement increases with strangeness content rather than with mass or baryon number of the hadron.
Such behaviour cannot be reproduced by any of the MC models commonly used,
suggesting that further developments are needed to obtain a complete microscopic understanding of strangeness production and indicating the presence of a phenomenon novel in high-multiplicity pp collisions.
The evolution of strangeness enhancement seen at the LHC steadily increases as a function of \average{\dnchdeta} from low multiplicity pp to high multiplicity \pPb and reaches the values observed in \PbPb collisions. This may point towards a common underlying physics mechanism which gradually compensates the strangeness suppression in fragmentation.
Further studies extending to higher multiplicity in small systems are essential, as they would demonstrate whether strangeness production saturates at the thermal equilibrium values predicted by the grand canonical statistical model~\cite{Cleymans:2006xj,Andronic:2008gu} or continues to increase. 
The remarkable similarity of strange particle production in pp, \pPb and \PbPb collisions adds to previous measurements in pp, which also exhibit characteristic features known from high-energy heavy-ion collisions~\cite{Khachatryan:2010gv,CMS:2012qk,Abelev:2012ola,Aad:2012gla,Aad:2013fja,Chatrchyan:2013nka,Abelev:2013haa,Adam:2015vsf,Khachatryan:2016yru,Khachatryan:2016txc} and are understood to be connected to the formation of a deconfined QCD phase at high temperature and energy density.

%% file: acknowledgements.tex

The ALICE Collaboration would like to thank all its engineers and technicians for their invaluable contributions to the construction of the experiment and the CERN accelerator teams for the outstanding performance of the LHC complex.
The ALICE Collaboration gratefully acknowledges the resources and support provided by all Grid centres and the Worldwide LHC Computing Grid (WLCG) collaboration.
The ALICE Collaboration acknowledges the following funding agencies for their support in building and
running the ALICE detector:
State Committee of Science,  World Federation of Scientists (WFS)
and Swiss Fonds Kidagan, Armenia;
Conselho Nacional de Desenvolvimento Cient\'{\i}fico e Tecnol\'{o}gico (CNPq), Financiadora de Estudos e Projetos (FINEP),
Funda\c{c}\~{a}o de Amparo \`{a} Pesquisa do Estado de S\~{a}o Paulo (FAPESP);
Ministry of Science \& Technology of China (MSTC), National Natural Science Foundation of China (NSFC) and Ministry of Education of China (MOEC)";
Ministry of Science, Education and Sports of Croatia and  Unity through Knowledge Fund, Croatia;
Ministry of Education and Youth of the Czech Republic;
Danish Natural Science Research Council, the Carlsberg Foundation and the Danish National Research Foundation;
The European Research Council under the European Community's Seventh Framework Programme;
Helsinki Institute of Physics and the Academy of Finland;
French CNRS-IN2P3, the `Region Pays de Loire', `Region Alsace', `Region Auvergne' and CEA, France;
German Bundesministerium fur Bildung, Wissenschaft, Forschung und Technologie (BMBF) and the Helmholtz Association;
General Secretariat for Research and Technology, Ministry of Development, Greece;
National Research, Development and Innovation Office (NKFIH), Hungary;
Council of Scientific and Industrial Research (CSIR), New Delhi;
Department of Atomic Energy and Department of Science and Technology of the Government of India;
Istituto Nazionale di Fisica Nucleare (INFN) and Centro Fermi - Museo Storico della Fisica e Centro Studi e Ricerche ``Enrico Fermi'', Italy;
Japan Society for the Promotion of Science (JSPS) KAKENHI and MEXT, Japan;
National Research Foundation of Korea (NRF);
Consejo Nacional de Cienca y Tecnologia (CONACYT), Direccion General de Asuntos del Personal Academico(DGAPA), M\'{e}xico, Amerique Latine Formation academique - 
European Commission~(ALFA-EC) and the EPLANET Program~(European Particle Physics Latin American Network);
Stichting voor Fundamenteel Onderzoek der Materie (FOM) and the Nederlandse Organisatie voor Wetenschappelijk Onderzoek (NWO), Netherlands;
Research Council of Norway (NFR);
Pontificia Universidad Cat\'{o}lica del Per\'{u};
National Science Centre, Poland;
Ministry of National Education/Institute for Atomic Physics and National Council of Scientific Research in Higher Education~(CNCSI-UEFISCDI), Romania;
Joint Institute for Nuclear Research, Dubna;
Ministry of Education and Science of Russian Federation, Russian Academy of Sciences, Russian Federal Agency of Atomic Energy, Russian Federal Agency for Science and Innovations and The Russian Foundation for Basic Research;
Ministry of Education of Slovakia;
Department of Science and Technology, South Africa;
Centro de Investigaciones Energeticas, Medioambientales y Tecnologicas (CIEMAT), E-Infrastructure shared between Europe and Latin America (EELA), 
Ministerio de Econom\'{i}a y Competitividad (MINECO) of Spain, Xunta de Galicia (Conseller\'{\i}a de Educaci\'{o}n),
Centro de Aplicaciones Tecnológicas y Desarrollo Nuclear (CEA\-DEN), Cubaenerg\'{\i}a, Cuba, and IAEA (International Atomic Energy Agency);
Swedish Research Council (VR) and Knut $\&$ Alice Wallenberg Foundation (KAW);
National Science and Technology Development Agency (NSDTA), Suranaree University of Technology (SUT) and Office of the Higher Education Commission under NRU project of Thailand;
Ukraine Ministry of Education and Science;
United Kingdom Science and Technology Facilities Council (STFC);
The United States Department of Energy, the United States National Science Foundation, the State of Texas, and the State of Ohio.

%% file: methods.tex

\section{Methods}

A detailed description of the ALICE detector and of its performance can be found in~\cite{Aamodt:2008zz,Abelev:2014ffa}. We briefly outline the main detectors utilized for this analysis.
The V0 detectors are two scintillator hodoscopes employed for triggering, background suppression and event-class determination. They are placed on either side of the interaction region at $z~=~3.3$~m and $z~=~-0.9$~m, covering the pseudorapidity regions $2.8~<~\eta~<~5.1$ and $-3.7~<~\eta~<~-1.7$, respectively.
Vertex reconstruction, central-barrel tracking and charged-hadron identification are performed with the Inner Tracking System (ITS) and the Time-Projection Chamber (TPC), which are located inside a solenoidal magnet providing a 0.5~T magnetic field.
The ITS is composed of six cylindrical layers of high-resolution silicon tracking detectors. The innermost layers consist of two arrays of hybrid Silicon Pixel Detectors (SPD) located at average radii 3.9 and 7.6~cm from the beam axis and covering \etaless{2.0} and \etaless{1.4}, respectively.
The TPC is a large cylindrical drift detector of radial and longitudinal size of about $85~<~r~<~250$~cm and $-250~<~z~<~250$~cm, respectively. It provides charged-hadron identification information via ionisation energy loss in the fill gas.

The data were collected in 2010 using a minimum-bias trigger requiring a hit in either the V0 scintillators or in the SPD detector, in coincidence with the arrival of proton bunches from both directions.
The contamination from beam-induced background is removed offline by using the timing information and correlations in the V0 and SPD detectors, as discussed in details in~\cite{Abelev:2014ffa}.
Events used for the data analysis are further required to have a reconstructed vertex within $\left|z\right|~<$~10~cm. Events containing more than one distinct vertex are tagged as pileup and are discarded. The remaining pileup fraction is estimated to be negligible, ranging from about $10^{-4}$ to $10^{-2}$ for the lowest and highest multiplicity classes, respectively. A total of about 100 million events has been utilised for the analysis.

The mean pseudorapidity densities of primary charged particles \average{\dnchdeta} are measured at midrapidity, \etaless{0.5}, for each event class using the technique described in~\cite{ALICE:2012xs}. The \average{\dnchdeta} values, corrected for acceptance and efficiency, as well as for contamination from secondary particles and combinatorial background, are listed in Table~\ref{tab:multi}. The relative RMS width of the corresponding multiplicity distributions ranges from 68\% to 30\% for the lowest and highest multiplicity classes, respectively. The corresponding fractions of the \inelgtzero cross-section are also summarized in Table~\ref{tab:multi}.

Strange \pKzero, \pLambda and \apLambda and multi-strange \pXi, \apXi, \pOmega and \apOmega candidates are reconstructed via topological selection criteria and invariant-mass analysis of their characteristic weak decays~\cite{Agashe:2014kda}:
\begin{center}
  \centering
\addtolength{\tabcolsep}{-3pt} 
  \begin{tabular*}{\linewidth}{rcll}
    \footnotesize \pKzero & $\rightarrow$ & \pPiplus + \pPiminus & {\footnotesize B.R. = (69.20 $\pm$ 0.05) \%} \\
    \footnotesize \pLambdaNS(\apLambdaNS) & $\rightarrow$ & \pProtonNS(\apProtonNS) + \pPiminusNS(\pPiplusNS) & {\footnotesize B.R. = (63.9 $\pm$ 0.5) \%} \\
    \footnotesize \pXiNS(\apXiNS) & $\rightarrow$ & \pLambdaNS(\apLambdaNS) + \pPiminusNS(\pPiplusNS) & {\footnotesize B.R. = (99.887 $\pm$ 0.035) \%} \\ 
    \footnotesize \pOmegaNS(\apOmegaNS) & $\rightarrow$ & \pLambdaNS(\apLambdaNS) + \pKminusNS(\pKplusNS) & {\footnotesize B.R. = (67.8 $\pm$ 0.7) \%} \\
  \end{tabular*}
\addtolength{\tabcolsep}{+3pt}
\end{center}
Details on the analysis technique are described in~\cite{Abelev:2013haa,Aamodt:2011zza,Abelev:2012jp}.
The results are corrected for detector acceptance and reconstruction efficiency calculated using events from the PYTHIA6 (tune Perugia~0) MC generator~\cite{Skands:2010ak} with particle transport performed via a GEANT3~\cite{Brun:1994aa} simulation of the ALICE detector. The contamination to \pLambda (\apLambda) yields from weak decays of charged and neutral \sXi baryons (feed-down) is subtracted using a data-driven approach~\cite{Abelev:2013haa}.
The study of systematic uncertainties follows the analysis described in~\cite{Abelev:2013haa,Aamodt:2011zza,Abelev:2012jp}.
Contributions common to all event classes (\nch-independent) are estimated and removed to determine the remaining uncertainties which are uncorrelated across different multiplicity intervals. The main sources of systematic uncertainty and their corresponding values are summarized in Table~\ref{tab:sys}.
The results on pion and proton production have been obtained following the analysis method discussed in~\cite{Adam:2015qaa}.


\begin{sidewaystable*}[p]
  \begin{minipage}{\linewidth}
    \centering
    \caption{Event multiplicity classes, their corresponding fraction of the \inelgtzero cross-section ($\sigma/\sigma_{\rm INEL>0}$) and their corresponding \average{\dnchdeta} at midrapidity (\etaless{0.5}). The value of \average{\dnchdeta} in the inclusive (\inelgtzero) class is 5.96 $\pm$ 0.23. The uncertainties are the quadratic sum of statistical and systematic contributions and represent standard deviations.
    }
    \label{tab:multi}
    \begin{tabularx}{\textwidth}{l*{10}{Y}}
      \toprule
      Class name & I & II & III & IV & V & VI & VII & VIII & IX & X \\
      \midrule
      $\sigma/\sigma_{\rm INEL>0}$ & 0--0.95\% & 0.95--4.7\% & 4.7--9.5\% & 9.5--14\% & 14--19\% & 19--28\% & 28--38\% & 38--48\% & 48--68\% & 68--100\% \\
      \average{\dnchdeta} & 21.3$\pm$0.6 & 16.5$\pm$0.5 & 13.5$\pm$0.4 & 11.5$\pm$0.3 & 10.1$\pm$0.3 & 8.45$\pm$0.25 & 6.72$\pm$0.21 & 5.40$\pm$0.17 & 3.90$\pm$0.14 & 2.26$\pm$0.12 \\
      \bottomrule
    \end{tabularx}
  \end{minipage}

  \vspace{2cm}
  
  \begin{minipage}{\linewidth}
    \caption{Main sources and values of the relative systematic uncertainties (standard deviations expressed in \%) of the \pt-differential yields. The values are reported for low, intermediate and high \pt. The sums of the contributions common to all event classes are listed separately as \nch-independent systematics.}
    \label{tab:sys}  
    \begin{tabularx}{\textwidth}{l*{3}{Y}*{3}{Y}*{3}{Y}*{3}{Y}}
      \toprule
      Hadron &  \multicolumn{3}{c}{\pKzero} & \multicolumn{3}{c}{\pLambda(\apLambda)} & \multicolumn{3}{c}{\pXi(\apXi)} & \multicolumn{3}{c}{\pOmega(\apOmega)} \\
      \pt (\GeVc)   		&   0.05 & 6.2 & 11.0   &   0.5 & 3.7 & 7.2    &   0.8 & 2.1 & 5.8   &   1.2 & 2.8 & 4.7 \\
      \cmidrule(lr){2-4} \cmidrule(lr){5-7} \cmidrule(lr){8-10} \cmidrule(lr){11-13} 
      Material budget 		&   4.0 & 4.0 & 4.0 	& 4.0 & 4.0 & 4.0 		&   4.0 & 4.0 & 4.0 	&   4.0 & 4.0 & 4.0 \\
      Transport code 		& \multicolumn{3}{c}{negligible}  & 1.0 & 1.0 & 1.0 	&   1.0 & 1.0 & 1.0 	&   1.0 & 1.0 & 1.0  \\
      Track selection  		&   1.0 & 5.0 & 0.8   	&   0.2 & 5.9 & 4.3   		&   0.4 & 0.3 & 2.2   	&   0.8 & 0.6 & 4.1  \\ 
      Topological selection   	&   2.6 & 1.1 & 2.3   	&   0.8 & 0.6 & 3.2   		&   3.1 & 2.0 & 4.0   	&   5.0 & 5.6 & 8.1 \\ 
      Particle identification   &   0.1 & 0.1 & 0.1   	&   0.2 & 0.2 & 3.0  		&   1.0 & 0.2 & 1.2   	&   1.1 & 1.7 & 3.2 \\ 
      Efficiency determination 	&   2.0 & 2.0 & 2.0  	&   2.0 & 2.0 & 2.0  		&   2.0 & 2.0 & 2.0   	&   2.0 & 2.0 & 2.0 \\ 
      Signal extraction   	&   1.5 & 1.2 & 3.6   	&   0.6 & 0.7 & 3.0   		&   1.5 & 0.2 & 1.0   	&   3.2 & 2.5 & 2.3 \\
      Proper lifetime   	&   1.3 & 0.1 & 0.2   	&   0.3 & 2.3 & 0.1    		&   0.9 & 0.1 & 0.1   	&   2.2 & 0.7 & 0.7 \\
      Competing decay rejection &   negl. & 0.7 & 1.3   &  negl. & 1.0 & 6.2    	&  \multicolumn{3}{c}{not applicable}  &  0.2 & 4.2 & 5.2  \\
      Feed-down correction 	& \multicolumn{3}{c}{not applicable}   &   3.3 & 2.1 & 4.3  & \multicolumn{3}{c}{negligible}  &  \multicolumn{3}{c}{negligible} \\
      \cmidrule(lr){2-4} \cmidrule(lr){5-7} \cmidrule(lr){8-10} \cmidrule(lr){11-13} 
      Total  			&   5.6 & 6.9 & 6.4   	&   5.8 & 8.2 & 11.2   	&  5.9 & 5.0 & 6.7   &   7.9 & 9.0 & 12.1 \\ 
      Common (\nch-independent) &   5.0 & 5.9 & 4.4   	&   5.4 & 7.8 & 9.9   	&   5.2 & 4.5 & 6.2   & 7.3 & 8.7 & 11.6  \\ 
      \bottomrule
    \end{tabularx}
  \end{minipage}
\end{sidewaystable*}

%% file: Alice_Authorlist_2016-May-30.tex


\begingroup
\small
\begin{flushleft}
J.~Adam\Irefn{org40}\And
D.~Adamov\'{a}\Irefn{org86}\And
M.M.~Aggarwal\Irefn{org90}\And
G.~Aglieri Rinella\Irefn{org36}\And
M.~Agnello\Irefn{org32}\textsuperscript{,}\Irefn{org112}\And
N.~Agrawal\Irefn{org49}\And
Z.~Ahammed\Irefn{org135}\And
S.~Ahmad\Irefn{org19}\And
S.U.~Ahn\Irefn{org70}\And
S.~Aiola\Irefn{org139}
% ...
\renewcommand\labelenumi{\textsuperscript{\theenumi}~}

\section*{Affiliation notes}
\renewcommand\theenumi{\roman{enumi}}
\begin{Authlist}
\item \Adef{0}Deceased
\item \Adef{idp1803184}{Also at: Georgia State University, Atlanta, Georgia, United States}
\item \Adef{idp3208048}{Also at: Also at Department of Applied Physics, Aligarh Muslim University, Aligarh, India}
\item \Adef{idp3929568}{Also at: M.V. Lomonosov Moscow State University, D.V. Skobeltsyn Institute of Nuclear, Physics, Moscow, Russia}
\end{Authlist}

\section*{Collaboration Institutes}
\renewcommand\theenumi{\arabic{enumi}~}
\begin{Authlist}

\item \Idef{org1}A.I. Alikhanyan National Science Laboratory (Yerevan Physics Institute) Foundation, Yerevan, Armenia
\item \Idef{org2}Benem\'{e}rita Universidad Aut\'{o}noma de Puebla, Puebla, Mexico
\item \Idef{org3}Bogolyubov Institute for Theoretical Physics, Kiev, Ukraine
\item \Idef{org4}Bose Institute, Department of Physics and Centre for Astroparticle Physics and Space Science (CAPSS), Kolkata, India
\item \Idef{org5}Budker Institute for Nuclear Physics, Novosibirsk, Russia
\item \Idef{org6}California Polytechnic State University, San Luis Obispo, California, United States
\item \Idef{org7}Central China Normal University, Wuhan, China
\item \Idef{org8}Centre de Calcul de l'IN2P3, Villeurbanne, France
\item \Idef{org9}Centro de Aplicaciones Tecnol\'{o}gicas y Desarrollo Nuclear (CEADEN), Havana, Cuba
\item \Idef{org10}Centro de Investigaciones Energ\'{e}ticas Medioambientales y Tecnol\'{o}gicas (CIEMAT), Madrid, Spain
\item \Idef{org11}Centro de Investigaci\'{o}n y de Estudios Avanzados (CINVESTAV), Mexico City and M\'{e}rida, Mexico
\item \Idef{org12}Centro Fermi - Museo Storico della Fisica e Centro Studi e Ricerche ``Enrico Fermi'', Rome, Italy
\item \Idef{org13}Chicago State University, Chicago, Illinois, USA
\item \Idef{org14}China Institute of Atomic Energy, Beijing, China
\item \Idef{org15}Commissariat \`{a} l'Energie Atomique, IRFU, Saclay, France
\item \Idef{org16}COMSATS Institute of Information Technology (CIIT), Islamabad, Pakistan
\item \Idef{org17}Departamento de F\'{\i}sica de Part\'{\i}culas and IGFAE, Universidad de Santiago de Compostela, Santiago de Compostela, Spain
\item \Idef{org18}Department of Physics and Technology, University of Bergen, Bergen, Norway
\item \Idef{org19}Department of Physics, Aligarh Muslim University, Aligarh, India
\item \Idef{org20}Department of Physics, Ohio State University, Columbus, Ohio, United States
\item \Idef{org21}Department of Physics, Sejong University, Seoul, South Korea
\item \Idef{org22}Department of Physics, University of Oslo, Oslo, Norway
\item \Idef{org23}Dipartimento di Fisica dell'Universit\`{a} 'La Sapienza' and Sezione INFN Rome, Italy
\item \Idef{org24}Dipartimento di Fisica dell'Universit\`{a} and Sezione INFN, Cagliari, Italy
\item \Idef{org25}Dipartimento di Fisica dell'Universit\`{a} and Sezione INFN, Trieste, Italy
\item \Idef{org26}Dipartimento di Fisica dell'Universit\`{a} and Sezione INFN, Turin, Italy
\item \Idef{org27}Dipartimento di Fisica e Astronomia dell'Universit\`{a} and Sezione INFN, Bologna, Italy
\item \Idef{org28}Dipartimento di Fisica e Astronomia dell'Universit\`{a} and Sezione INFN, Catania, Italy
\item \Idef{org29}Dipartimento di Fisica e Astronomia dell'Universit\`{a} and Sezione INFN, Padova, Italy
\item \Idef{org30}Dipartimento di Fisica `E.R.~Caianiello' dell'Universit\`{a} and Gruppo Collegato INFN, Salerno, Italy
\item \Idef{org31}Dipartimento di Scienze e Innovazione Tecnologica dell'Universit\`{a} del  Piemonte Orientale and Gruppo Collegato INFN, Alessandria, Italy
\item \Idef{org32}Dipartimento DISAT del Politecnico and Sezione INFN, Turin, Italy
\item \Idef{org33}Dipartimento Interateneo di Fisica `M.~Merlin' and Sezione INFN, Bari, Italy
\item \Idef{org34}Division of Experimental High Energy Physics, University of Lund, Lund, Sweden
\item \Idef{org35}Eberhard Karls Universit\"{a}t T\"{u}bingen, T\"{u}bingen, Germany
\item \Idef{org36}European Organization for Nuclear Research (CERN), Geneva, Switzerland
\item \Idef{org37}Excellence Cluster Universe, Technische Universit\"{a}t M\"{u}nchen, Munich, Germany
\item \Idef{org38}Faculty of Engineering, Bergen University College, Bergen, Norway
\item \Idef{org39}Faculty of Mathematics, Physics and Informatics, Comenius University, Bratislava, Slovakia
\item \Idef{org40}Faculty of Nuclear Sciences and Physical Engineering, Czech Technical University in Prague, Prague, Czech Republic
\item \Idef{org41}Faculty of Science, P.J.~\v{S}af\'{a}rik University, Ko\v{s}ice, Slovakia
\item \Idef{org42}Faculty of Technology, Buskerud and Vestfold University College, Vestfold, Norway
\item \Idef{org43}Frankfurt Institute for Advanced Studies, Johann Wolfgang Goethe-Universit\"{a}t Frankfurt, Frankfurt, Germany
\item \Idef{org44}Gangneung-Wonju National University, Gangneung, South Korea
\item \Idef{org45}Gauhati University, Department of Physics, Guwahati, India
\item \Idef{org46}Helmholtz-Institut f\"{u}r Strahlen- und Kernphysik, Rheinische Friedrich-Wilhelms-Universit\"{a}t Bonn, Bonn, Germany
\item \Idef{org47}Helsinki Institute of Physics (HIP), Helsinki, Finland
\item \Idef{org48}Hiroshima University, Hiroshima, Japan
\item \Idef{org49}Indian Institute of Technology Bombay (IIT), Mumbai, India
\item \Idef{org50}Indian Institute of Technology Indore, Indore (IITI), India
\item \Idef{org51}Indonesian Institute of Sciences, Jakarta, Indonesia
\item \Idef{org52}Inha University, Incheon, South Korea
\item \Idef{org53}Institut de Physique Nucl\'eaire d'Orsay (IPNO), Universit\'e Paris-Sud, CNRS-IN2P3, Orsay, France
\item \Idef{org54}Institut f\"{u}r Informatik, Johann Wolfgang Goethe-Universit\"{a}t Frankfurt, Frankfurt, Germany
\item \Idef{org55}Institut f\"{u}r Kernphysik, Johann Wolfgang Goethe-Universit\"{a}t Frankfurt, Frankfurt, Germany
\item \Idef{org56}Institut f\"{u}r Kernphysik, Westf\"{a}lische Wilhelms-Universit\"{a}t M\"{u}nster, M\"{u}nster, Germany
\item \Idef{org57}Institut Pluridisciplinaire Hubert Curien (IPHC), Universit\'{e} de Strasbourg, CNRS-IN2P3, Strasbourg, France
\item \Idef{org58}Institute for Nuclear Research, Academy of Sciences, Moscow, Russia
\item \Idef{org59}Institute for Subatomic Physics of Utrecht University, Utrecht, Netherlands
\item \Idef{org60}Institute for Theoretical and Experimental Physics, Moscow, Russia
\item \Idef{org61}Institute of Experimental Physics, Slovak Academy of Sciences, Ko\v{s}ice, Slovakia
\item \Idef{org62}Institute of Physics, Academy of Sciences of the Czech Republic, Prague, Czech Republic
\item \Idef{org63}Institute of Physics, Bhubaneswar, India
\item \Idef{org64}Institute of Space Science (ISS), Bucharest, Romania
\item \Idef{org65}Instituto de Ciencias Nucleares, Universidad Nacional Aut\'{o}noma de M\'{e}xico, Mexico City, Mexico
\item \Idef{org66}Instituto de F\'{\i}sica, Universidad Nacional Aut\'{o}noma de M\'{e}xico, Mexico City, Mexico
\item \Idef{org67}iThemba LABS, National Research Foundation, Somerset West, South Africa
\item \Idef{org68}Joint Institute for Nuclear Research (JINR), Dubna, Russia
\item \Idef{org69}Konkuk University, Seoul, South Korea
\item \Idef{org70}Korea Institute of Science and Technology Information, Daejeon, South Korea
\item \Idef{org71}KTO Karatay University, Konya, Turkey
\item \Idef{org72}Laboratoire de Physique Corpusculaire (LPC), Clermont Universit\'{e}, Universit\'{e} Blaise Pascal, CNRS--IN2P3, Clermont-Ferrand, France
\item \Idef{org73}Laboratoire de Physique Subatomique et de Cosmologie, Universit\'{e} Grenoble-Alpes, CNRS-IN2P3, Grenoble, France
\item \Idef{org74}Laboratori Nazionali di Frascati, INFN, Frascati, Italy
\item \Idef{org75}Laboratori Nazionali di Legnaro, INFN, Legnaro, Italy
\item \Idef{org76}Lawrence Berkeley National Laboratory, Berkeley, California, United States
\item \Idef{org77}Moscow Engineering Physics Institute, Moscow, Russia
\item \Idef{org78}Nagasaki Institute of Applied Science, Nagasaki, Japan
\item \Idef{org79}National Centre for Nuclear Studies, Warsaw, Poland
\item \Idef{org80}National Institute for Physics and Nuclear Engineering, Bucharest, Romania
\item \Idef{org81}National Institute of Science Education and Research, Bhubaneswar, India
\item \Idef{org82}National Research Centre Kurchatov Institute, Moscow, Russia
\item \Idef{org83}Niels Bohr Institute, University of Copenhagen, Copenhagen, Denmark
\item \Idef{org84}Nikhef, Nationaal instituut voor subatomaire fysica, Amsterdam, Netherlands
\item \Idef{org85}Nuclear Physics Group, STFC Daresbury Laboratory, Daresbury, United Kingdom
\item \Idef{org86}Nuclear Physics Institute, Academy of Sciences of the Czech Republic, \v{R}e\v{z} u Prahy, Czech Republic
\item \Idef{org87}Oak Ridge National Laboratory, Oak Ridge, Tennessee, United States
\item \Idef{org88}Petersburg Nuclear Physics Institute, Gatchina, Russia
\item \Idef{org89}Physics Department, Creighton University, Omaha, Nebraska, United States
\item \Idef{org90}Physics Department, Panjab University, Chandigarh, India
\item \Idef{org91}Physics Department, University of Athens, Athens, Greece
\item \Idef{org92}Physics Department, University of Cape Town, Cape Town, South Africa
\item \Idef{org93}Physics Department, University of Jammu, Jammu, India
\item \Idef{org94}Physics Department, University of Rajasthan, Jaipur, India
\item \Idef{org95}Physik Department, Technische Universit\"{a}t M\"{u}nchen, Munich, Germany
\item \Idef{org96}Physikalisches Institut, Ruprecht-Karls-Universit\"{a}t Heidelberg, Heidelberg, Germany
\item \Idef{org97}Purdue University, West Lafayette, Indiana, United States
\item \Idef{org98}Pusan National University, Pusan, South Korea
\item \Idef{org99}Research Division and ExtreMe Matter Institute EMMI, GSI Helmholtzzentrum f\"ur Schwerionenforschung, Darmstadt, Germany
\item \Idef{org100}Rudjer Bo\v{s}kovi\'{c} Institute, Zagreb, Croatia
\item \Idef{org101}Russian Federal Nuclear Center (VNIIEF), Sarov, Russia
\item \Idef{org102}Saha Institute of Nuclear Physics, Kolkata, India
\item \Idef{org103}School of Physics and Astronomy, University of Birmingham, Birmingham, United Kingdom
\item \Idef{org104}Secci\'{o}n F\'{\i}sica, Departamento de Ciencias, Pontificia Universidad Cat\'{o}lica del Per\'{u}, Lima, Peru
\item \Idef{org105}Sezione INFN, Bari, Italy
\item \Idef{org106}Sezione INFN, Bologna, Italy
\item \Idef{org107}Sezione INFN, Cagliari, Italy
\item \Idef{org108}Sezione INFN, Catania, Italy
\item \Idef{org109}Sezione INFN, Padova, Italy
\item \Idef{org110}Sezione INFN, Rome, Italy
\item \Idef{org111}Sezione INFN, Trieste, Italy
\item \Idef{org112}Sezione INFN, Turin, Italy
\item \Idef{org113}SSC IHEP of NRC Kurchatov institute, Protvino, Russia
\item \Idef{org114}Stefan Meyer Institut f\"{u}r Subatomare Physik (SMI), Vienna, Austria
\item \Idef{org115}SUBATECH, Ecole des Mines de Nantes, Universit\'{e} de Nantes, CNRS-IN2P3, Nantes, France
\item \Idef{org116}Suranaree University of Technology, Nakhon Ratchasima, Thailand
\item \Idef{org117}Technical University of Ko\v{s}ice, Ko\v{s}ice, Slovakia
\item \Idef{org118}Technical University of Split FESB, Split, Croatia
\item \Idef{org119}The Henryk Niewodniczanski Institute of Nuclear Physics, Polish Academy of Sciences, Cracow, Poland
\item \Idef{org120}The University of Texas at Austin, Physics Department, Austin, Texas, USA
\item \Idef{org121}Universidad Aut\'{o}noma de Sinaloa, Culiac\'{a}n, Mexico
\item \Idef{org122}Universidade de S\~{a}o Paulo (USP), S\~{a}o Paulo, Brazil
\item \Idef{org123}Universidade Estadual de Campinas (UNICAMP), Campinas, Brazil
\item \Idef{org124}University of Houston, Houston, Texas, United States
\item \Idef{org125}University of Jyv\"{a}skyl\"{a}, Jyv\"{a}skyl\"{a}, Finland
\item \Idef{org126}University of Liverpool, Liverpool, United Kingdom
\item \Idef{org127}University of Tennessee, Knoxville, Tennessee, United States
\item \Idef{org128}University of the Witwatersrand, Johannesburg, South Africa
\item \Idef{org129}University of Tokyo, Tokyo, Japan
\item \Idef{org130}University of Tsukuba, Tsukuba, Japan
\item \Idef{org131}University of Zagreb, Zagreb, Croatia
\item \Idef{org132}Universit\'{e} de Lyon, Universit\'{e} Lyon 1, CNRS/IN2P3, IPN-Lyon, Villeurbanne, France
\item \Idef{org133}Universit\`{a} di Brescia, Brescia, Italy
\item \Idef{org134}V.~Fock Institute for Physics, St. Petersburg State University, St. Petersburg, Russia
\item \Idef{org135}Variable Energy Cyclotron Centre, Kolkata, India
\item \Idef{org136}Warsaw University of Technology, Warsaw, Poland
\item \Idef{org137}Wayne State University, Detroit, Michigan, United States
\item \Idef{org138}Wigner Research Centre for Physics, Hungarian Academy of Sciences, Budapest, Hungary
\item \Idef{org139}Yale University, New Haven, Connecticut, United States
\item \Idef{org140}Yonsei University, Seoul, South Korea
\item \Idef{org141}Zentrum f\"{u}r Technologietransfer und Telekommunikation (ZTT), Fachhochschule Worms, Worms, Germany
\end{Authlist}
\endgroup

%% file: cds.bbl
\providecommand{\href}[2]{#2}\begingroup\raggedright\begin{thebibliography}{10}

\bibitem{Shuryak:1980tp}
E.~V. Shuryak, ``{Quantum Chromodynamics and the Theory of Superdense
  Matter},''
\href{http://dx.doi.org/10.1016/0370-1573(80)90105-2}{{\em Phys. Rept.}
  {\bfseries 61} (1980) 71--158}.

\bibitem{Andersen:1999ym}
{\bfseries WA97} Collaboration, E.~Andersen {\em et~al.}, ``{Strangeness
  enhancement at mid-rapidity in Pb--Pb collisions at 158 AGeV/c},''
\href{http://dx.doi.org/10.1016/S0370-2693(99)00140-9}{{\em Phys. Lett.}
  {\bfseries B449} (1999) 401--406}.

\bibitem{Afanasiev:2002he}
{\bfseries NA49} Collaboration, S.~V. Afanasiev {\em et~al.}, ``{$\Xi$ and
  $\bar{\Xi}$ production in central Pb+Pb collisions at 158 GeV/c per
  nucleon},'' \href{http://dx.doi.org/10.1016/S0370-2693(02)01970-6}{{\em Phys.
  Lett.} {\bfseries B538} (2002) 275--281},
\href{http://arxiv.org/abs/hep-ex/0202037}{{\ttfamily arXiv:hep-ex/0202037
  [hep-ex]}}.

\bibitem{Antinori:2004ee}
{\bfseries NA57} Collaboration, F.~Antinori {\em et~al.}, ``{Energy dependence
  of hyperon production in nucleus nucleus collisions at SPS},''
  \href{http://dx.doi.org/10.1016/j.physletb.2004.05.025}{{\em Phys. Lett.}
  {\bfseries B595} (2004) 68--74},
\href{http://arxiv.org/abs/nucl-ex/0403022}{{\ttfamily arXiv:nucl-ex/0403022
  [nucl-ex]}}.

\bibitem{Abelev:2007xp}
{\bfseries STAR} Collaboration, B.~I. Abelev {\em et~al.}, ``{Enhanced strange
  baryon production in Au+Au collisions compared to p+p at $\sqrt{s}$ =
  200-GeV},'' \href{http://dx.doi.org/10.1103/PhysRevC.77.044908}{{\em Phys.
  Rev.} {\bfseries C77} (2008) 044908},
\href{http://arxiv.org/abs/0705.2511}{{\ttfamily arXiv:0705.2511 [nucl-ex]}}.

\bibitem{ABELEV:2013zaa}
{\bfseries ALICE} Collaboration, B.~Abelev {\em et~al.}, ``{Multi-strange
  baryon production at mid-rapidity in Pb-Pb collisions at $\sqrt{s_{\rm NN}}$
  = 2.76 TeV},'' \href{http://dx.doi.org/10.1016/j.physletb.2014.05.052,
  10.1016/j.physletb.2013.11.048}{{\em Phys. Lett.} {\bfseries B728} (2014)
  216--227}, \href{http://arxiv.org/abs/1307.5543}{{\ttfamily arXiv:1307.5543
  [nucl-ex]}}.
[Erratum: Phys. Lett. B734, 409 (2014)].

\bibitem{Koch:1986ud}
P.~Koch, B.~Muller, and J.~Rafelski, ``{Strangeness in Relativistic Heavy Ion
  Collisions},''
\href{http://dx.doi.org/10.1016/0370-1573(86)90096-7}{{\em Phys. Rept.}
  {\bfseries 142} (1986) 167--262}.

\bibitem{Khachatryan:2010gv}
{\bfseries CMS} Collaboration, V.~Khachatryan {\em et~al.}, ``{Observation of
  Long-Range Near-Side Angular Correlations in Proton-Proton Collisions at the
  LHC},'' \href{http://dx.doi.org/10.1007/JHEP09(2010)091}{{\em JHEP}
  {\bfseries 09} (2010) 091},
\href{http://arxiv.org/abs/1009.4122}{{\ttfamily arXiv:1009.4122 [hep-ex]}}.

\bibitem{Khachatryan:2016txc}
{\bfseries CMS} Collaboration, V.~Khachatryan {\em et~al.}, ``{Evidence for
  collectivity in pp collisions at the LHC},''
  \href{http://dx.doi.org/10.1016/j.physletb.2016.12.009}{{\em Phys. Lett.}
  {\bfseries B765} (2017) 193--220},
\href{http://arxiv.org/abs/1606.06198}{{\ttfamily arXiv:1606.06198 [nucl-ex]}}.

\bibitem{Abelev:2013haa}
{\bfseries ALICE} Collaboration, B.~Abelev {\em et~al.}, ``{Multiplicity
  Dependence of Pion, Kaon, Proton and Lambda Production in p--Pb Collisions at
  $\sqrt{s_{\rm NN}}$ = 5.02 TeV},''
  \href{http://dx.doi.org/10.1016/j.physletb.2013.11.020}{{\em Phys. Lett.}
  {\bfseries B728} (2014) 25--38},
\href{http://arxiv.org/abs/1307.6796}{{\ttfamily arXiv:1307.6796 [nucl-ex]}}.

\bibitem{Adam:2015vsf}
{\bfseries ALICE} Collaboration, J.~Adam {\em et~al.}, ``{Multi-strange baryon
  production in p-Pb collisions at $\sqrt{s_\mathbf{NN}}=5.02$ TeV},''
  \href{http://dx.doi.org/10.1016/j.physletb.2016.05.027}{{\em Phys. Lett.}
  {\bfseries B758} (2016) 389--401},
\href{http://arxiv.org/abs/1512.07227}{{\ttfamily arXiv:1512.07227 [nucl-ex]}}.

\bibitem{Cleymans:2006xj}
J.~Cleymans, I.~Kraus, H.~Oeschler, K.~Redlich, and S.~Wheaton, ``{Statistical
  model predictions for particle ratios at $\sqrt{s_{\rm NN}}$ = 5.5-TeV},''
  \href{http://dx.doi.org/10.1103/PhysRevC.74.034903}{{\em Phys. Rev.}
  {\bfseries C74} (2006) 034903},
\href{http://arxiv.org/abs/hep-ph/0604237}{{\ttfamily arXiv:hep-ph/0604237
  [hep-ph]}}.

\bibitem{Andronic:2008gu}
A.~Andronic, P.~Braun-Munzinger, and J.~Stachel, ``{Thermal hadron production
  in relativistic nuclear collisions: The Hadron mass spectrum, the horn, and
  the QCD phase transition},''
  \href{http://dx.doi.org/10.1016/j.physletb.2009.02.014,
  10.1016/j.physletb.2009.06.021}{{\em Phys. Lett.} {\bfseries B673} (2009)
  142--145}, \href{http://arxiv.org/abs/0812.1186}{{\ttfamily arXiv:0812.1186
  [nucl-th]}}.
[Erratum: Phys. Lett.B678,516(2009)].

\bibitem{Hagedorn:1967ua}
R.~Hagedorn and J.~Ranft, ``{Statistical thermodynamics of strong interactions
  at high-energies. 2. Momentum spectra of particles produced in
  pp-collisions},''
{\em Nuovo Cim. Suppl.} {\bfseries 6} (1968) 169--354.

\bibitem{Becattini:1997rv}
F.~Becattini and U.~W. Heinz, ``{Thermal hadron production in p p and p anti-p
  collisions},'' \href{http://dx.doi.org/10.1007/s002880050551}{{\em Z. Phys.}
  {\bfseries C76} (1997) 269--286},
  \href{http://arxiv.org/abs/hep-ph/9702274}{{\ttfamily arXiv:hep-ph/9702274
  [hep-ph]}}.
[Erratum: Z. Phys.C76,578(1997)].

\bibitem{Redlich:2001kb}
K.~Redlich and A.~Tounsi, ``{Strangeness enhancement and energy dependence in
  heavy ion collisions},''
  \href{http://dx.doi.org/10.1007/s10052-002-0983-1}{{\em Eur. Phys. J.}
  {\bfseries C24} (2002) 589--594},
\href{http://arxiv.org/abs/hep-ph/0111261}{{\ttfamily arXiv:hep-ph/0111261
  [hep-ph]}}.

\bibitem{Becattini:2008yn}
F.~Becattini and J.~Manninen, ``{Strangeness production from SPS to LHC},''
  \href{http://dx.doi.org/10.1088/0954-3899/35/10/104013}{{\em J. Phys.}
  {\bfseries G35} (2008) 104013},
\href{http://arxiv.org/abs/0805.0098}{{\ttfamily arXiv:0805.0098 [nucl-th]}}.

\bibitem{Aichelin:2008mi}
J.~Aichelin and K.~Werner, ``{Centrality Dependence of Strangeness Enhancement
  in Ultrarelativistic Heavy Ion Collisions: A Core-Corona Effect},''
  \href{http://dx.doi.org/10.1103/PhysRevC.79.064907,
  10.1103/PhysRevC.81.029902}{{\em Phys. Rev.} {\bfseries C79} (2009) 064907},
  \href{http://arxiv.org/abs/0810.4465}{{\ttfamily arXiv:0810.4465 [nucl-th]}}.
[Erratum: Phys. Rev.C81,029902(2010)].

\bibitem{CMS:2012qk}
{\bfseries CMS} Collaboration, S.~Chatrchyan {\em et~al.}, ``{Observation of
  long-range near-side angular correlations in proton-lead collisions at the
  LHC},'' \href{http://dx.doi.org/10.1016/j.physletb.2012.11.025}{{\em Phys.
  Lett.} {\bfseries B718} (2013) 795--814},
\href{http://arxiv.org/abs/1210.5482}{{\ttfamily arXiv:1210.5482 [nucl-ex]}}.

\bibitem{Abelev:2012ola}
{\bfseries ALICE} Collaboration, B.~Abelev {\em et~al.}, ``{Long-range angular
  correlations on the near and away side in p--Pb collisions at $\sqrt{s_{\rm
  NN}}=5.02$ TeV},''
  \href{http://dx.doi.org/10.1016/j.physletb.2013.01.012}{{\em Phys. Lett.}
  {\bfseries B719} (2013) 29--41},
\href{http://arxiv.org/abs/1212.2001}{{\ttfamily arXiv:1212.2001 [nucl-ex]}}.

\bibitem{Aad:2012gla}
{\bfseries ATLAS} Collaboration, G.~Aad {\em et~al.}, ``{Observation of
  Associated Near-Side and Away-Side Long-Range Correlations in $\sqrt{s_{\rm
  NN}}$=5.02 TeV Proton-Lead Collisions with the ATLAS Detector},''
  \href{http://dx.doi.org/10.1103/PhysRevLett.110.182302}{{\em Phys. Rev.
  Lett.} {\bfseries 110} no.~18, (2013) 182302},
\href{http://arxiv.org/abs/1212.5198}{{\ttfamily arXiv:1212.5198 [hep-ex]}}.

\bibitem{Aad:2013fja}
{\bfseries ATLAS} Collaboration, G.~Aad {\em et~al.}, ``{Measurement with the
  ATLAS detector of multi-particle azimuthal correlations in p+Pb collisions at
  $\sqrt{s_{\rm NN}}$=5.02 TeV},''
  \href{http://dx.doi.org/10.1016/j.physletb.2013.06.057}{{\em Phys. Lett.}
  {\bfseries B725} (2013) 60--78},
\href{http://arxiv.org/abs/1303.2084}{{\ttfamily arXiv:1303.2084 [hep-ex]}}.

\bibitem{Chatrchyan:2013nka}
{\bfseries CMS} Collaboration, S.~Chatrchyan {\em et~al.}, ``{Multiplicity and
  transverse momentum dependence of two- and four-particle correlations in pPb
  and PbPb collisions},''
  \href{http://dx.doi.org/10.1016/j.physletb.2013.06.028}{{\em Phys. Lett.}
  {\bfseries B724} (2013) 213--240},
\href{http://arxiv.org/abs/1305.0609}{{\ttfamily arXiv:1305.0609 [nucl-ex]}}.

\bibitem{ABELEV:2013wsa}
{\bfseries ALICE} Collaboration, B.~B. Abelev {\em et~al.}, ``{Long-range
  angular correlations of $\rm \pi$, K and p in p-Pb collisions at
  $\sqrt{s_{\rm NN}}$ = 5.02 TeV},''
  \href{http://dx.doi.org/10.1016/j.physletb.2013.08.024}{{\em Phys. Lett.}
  {\bfseries B726} (2013) 164--177},
\href{http://arxiv.org/abs/1307.3237}{{\ttfamily arXiv:1307.3237 [nucl-ex]}}.

\bibitem{Khachatryan:2016yru}
{\bfseries CMS} Collaboration, V.~Khachatryan {\em et~al.}, ``{Multiplicity and
  rapidity dependence of strange hadron production in pp, pPb, and PbPb
  collisions at the LHC},''
  \href{http://dx.doi.org/10.1016/j.physletb.2017.01.075}{{\em Phys. Lett.}
  {\bfseries B768} (2017) 103--129},
\href{http://arxiv.org/abs/1605.06699}{{\ttfamily arXiv:1605.06699 [nucl-ex]}}.

\bibitem{Aamodt:2008zz}
{\bfseries ALICE} Collaboration, K.~Aamodt {\em et~al.}, ``{The ALICE
  experiment at the CERN LHC},''
\href{http://dx.doi.org/10.1088/1748-0221/3/08/S08002}{{\em JINST} {\bfseries
  3} (2008) S08002}.

\bibitem{Abelev:2013xaa}
{\bfseries ALICE} Collaboration, B.~Abelev {\em et~al.}, ``{$K^0_S$ and
  $\Lambda$ production in Pb--Pb collisions at $\sqrt{s_{\rm NN}}$ = 2.76
  TeV},'' \href{http://dx.doi.org/10.1103/PhysRevLett.111.222301}{{\em Phys.
  Rev. Lett.} {\bfseries 111} (2013) 222301},
\href{http://arxiv.org/abs/1307.5530}{{\ttfamily arXiv:1307.5530 [nucl-ex]}}.

\bibitem{Heinz:2004qz}
U.~W. Heinz, ``{Concepts of heavy ion physics},'' in {\em {2002 European School
  of high-energy physics, Pylos, Greece, 25 Aug-7 Sep 2002: Proceedings}},
  pp.~165--238.
\newblock 2004.
\newblock \href{http://arxiv.org/abs/hep-ph/0407360}{{\ttfamily
  arXiv:hep-ph/0407360 [hep-ph]}}.
\newblock
\url{http://doc.cern.ch/yellowrep/CERN-2004-001}.
\newblock

\bibitem{Schnedermann:1993ws}
E.~Schnedermann, J.~Sollfrank, and U.~W. Heinz, ``{Thermal phenomenology of
  hadrons from 200 A/GeV S+S collisions},''
  \href{http://dx.doi.org/10.1103/PhysRevC.48.2462}{{\em Phys. Rev.} {\bfseries
  C48} (1993) 2462--2475},
\href{http://arxiv.org/abs/nucl-th/9307020}{{\ttfamily arXiv:nucl-th/9307020
  [nucl-th]}}.

\bibitem{Sjostrand:2007gs}
T.~Sj{\"o}strand, S.~Mrenna, and P.~Z. Skands, ``{A Brief Introduction to
  PYTHIA 8.1},'' \href{http://dx.doi.org/10.1016/j.cpc.2008.01.036}{{\em
  Comput. Phys. Commun.} {\bfseries 178} (2008) 852--867},
\href{http://arxiv.org/abs/0710.3820}{{\ttfamily arXiv:0710.3820 [hep-ph]}}.

\bibitem{Pierog:2013ria}
T.~Pierog, I.~Karpenko, J.~M. Katzy, E.~Yatsenko, and K.~Werner, ``{EPOS LHC:
  Test of collective hadronization with data measured at the CERN Large Hadron
  Collider},'' \href{http://dx.doi.org/10.1103/PhysRevC.92.034906}{{\em Phys.
  Rev.} {\bfseries C92} no.~3, (2015) 034906},
\href{http://arxiv.org/abs/1306.0121}{{\ttfamily arXiv:1306.0121 [hep-ph]}}.

\bibitem{Bierlich:2015rha}
C.~Bierlich and J.~R. Christiansen, ``{Effects of Colour Reconnection on Hadron
  Flavour Observables},''
  \href{http://dx.doi.org/10.1103/PhysRevD.92.094010}{{\em Phys. Rev.}
  {\bfseries D92} (2015) 094010},
\href{http://arxiv.org/abs/1507.02091}{{\ttfamily arXiv:1507.02091 [hep-ph]}}.

\bibitem{Abelev:2014ffa}
{\bfseries ALICE} Collaboration, B.~Abelev {\em et~al.}, ``{Performance of the
  ALICE Experiment at the CERN LHC},''
  \href{http://dx.doi.org/10.1142/S0217751X14300440}{{\em Int. J. Mod. Phys.}
  {\bfseries A29} (2014) 1430044},
\href{http://arxiv.org/abs/1402.4476}{{\ttfamily arXiv:1402.4476 [nucl-ex]}}.

\bibitem{ALICE:2012xs}
{\bfseries ALICE} Collaboration, B.~Abelev {\em et~al.}, ``{Pseudorapidity
  density of charged particles in p+Pb collisions at $\sqrt{s_{\rm
  NN}}~=~5.02$~TeV},''
  \href{http://dx.doi.org/10.1103/PhysRevLett.110.032301}{{\em Phys. Rev.
  Lett.} {\bfseries 110} no.~3, (2013) 032301},
\href{http://arxiv.org/abs/1210.3615}{{\ttfamily arXiv:1210.3615 [nucl-ex]}}.

\bibitem{Agashe:2014kda}
{\bfseries Particle Data Group} Collaboration, K.~A. Olive {\em et~al.},
  ``{Review of Particle Physics},''
\href{http://dx.doi.org/10.1088/1674-1137/38/9/090001}{{\em Chin. Phys.}
  {\bfseries C38} (2014) 090001}.

\bibitem{Aamodt:2011zza}
{\bfseries ALICE} Collaboration, K.~Aamodt {\em et~al.}, ``{Strange particle
  production in proton-proton collisions at $\sqrt{s}$ = 0.9 TeV with ALICE at
  the LHC},'' \href{http://dx.doi.org/10.1140/epjc/s10052-011-1594-5}{{\em Eur.
  Phys. J.} {\bfseries C71} (2011) 1594},
\href{http://arxiv.org/abs/1012.3257}{{\ttfamily arXiv:1012.3257 [hep-ex]}}.

\bibitem{Abelev:2012jp}
{\bfseries ALICE} Collaboration, B.~Abelev {\em et~al.}, ``{Multi-strange
  baryon production in pp collisions at $\sqrt{s} = 7$ TeV with {ALICE}},''
  \href{http://dx.doi.org/10.1016/j.physletb.2012.05.011}{{\em Phys. Lett. B}
  {\bfseries 712} (2012) 309--318},
\href{http://arxiv.org/abs/1204.0282}{{\ttfamily arXiv:1204.0282 [nucl-ex]}}.

\bibitem{Skands:2010ak}
P.~Z. Skands, ``{Tuning Monte Carlo Generators: The Perugia Tunes},''
  \href{http://dx.doi.org/10.1103/PhysRevD.82.074018}{{\em Phys. Rev.}
  {\bfseries D82} (2010) 074018},
\href{http://arxiv.org/abs/1005.3457}{{\ttfamily arXiv:1005.3457 [hep-ph]}}.

\bibitem{Brun:1994aa}
R.~Brun, F.~Bruyant, F.~Carminati, S.~Giani, M.~Maire, A.~McPherson,
  G.~Patrick, and L.~Urban, ``{GEANT Detector Description and Simulation
  Tool},''
\href{http://arxiv.org/abs/CERN-W5013, CERN-W-5013, W5013, W-5013}{{\ttfamily
  CERN-W5013, CERN-W-5013, W5013, W-5013}}.

\bibitem{Adam:2015qaa}
{\bfseries ALICE} Collaboration, J.~Adam {\em et~al.}, ``{Measurement of pion,
  kaon and proton production in proton-proton collisions at $\sqrt{s} = 7$
  TeV},'' \href{http://dx.doi.org/10.1140/epjc/s10052-015-3422-9}{{\em Eur.
  Phys. J. C} {\bfseries 75} no.~5, (2015) 226},
\href{http://arxiv.org/abs/1504.00024}{{\ttfamily arXiv:1504.00024 [nucl-ex]}}.

\end{thebibliography}\endgroup
